\documentclass[aps,prb,twocolumn,showpacs,tightenlines,amsmath]{revtex4}
 \usepackage{graphics}
 \usepackage{epsfig}
   \begin{document}
   \title{\Large \bf{
Multi-particle correlations of an oscillating scatterer
                    } 
         }
\author{
M. Moskalets$^{1,2}$
and
M. B\"uttiker$^{1}$
}
\affiliation{
     $^1$D\'epartement de Physique Th\'eorique, Universit\'e de Gen\`eve,
     CH-1211 Gen\`eve 4, Switzerland\\
     $^2$Department of Metal and Semiconductor Physics,\\
     National Technical University "Kharkiv Polytechnic Institute",
     61002 Kharkiv, Ukraine\\}
\date\today
   \begin{abstract}
Using multi-particle distribution functions we calculate the correlations produced by a periodically driven scatterer in a system of noninteracting electrons at zero temperature.
The multi-particle correlations due to a quantum exchange effect are expressed in terms of photon-assisted scattering amplitudes.
The results we obtain are valid for slow but arbitrary in strength driving.
We show that even for large amplitude pumps the zero-frequency noise power is related to two-particle correlations.
In addition to two-particle correlations a large amplitude pump can generate multi-particle correlations.
\end{abstract}
\maketitle
\small

\section{Introduction}
\label{intro}

In recent papers
\cite{SamuelssonB04,BTT05}
it was shown that a periodically driven unbiased scatterer,
a quantum pump \cite{Brouwer98,SMCG99,PB03,AEGSS04,Vavilov05},
can generate entanglement in a system of noninteracting fermions.
Entanglement
\cite{Schrodinger35,EPR35,Bohr35,Bell65}
is a property of quantum objects which has a powerful potential for applications in information processing. \cite{BD00}
The main resource for quantum computation are specific correlations between different parts of a quantum system.
These correlations reflect the fact that some quantum states of a whole system can not be represented as a product of states corresponding to its different parts.
Such states are called non-separable.
For them the joint probability to find the system's parts in some states does not factorize into the product of probabilities to find any of the parts in a corresponding state.
Thus the joint probability consists of two terms: the first one is a product of probabilities for single parts, and the second one depends on correlations between the system's parts.
The last term in itself can be used as an entanglement measure (see, e.g., Refs.~\onlinecite{Cirone05, Kaplan05}).
Such a measure is especially appropriate for systems of non-interacting particles where there are no other sources of correlations except the quantum-statistical interactions due to the Pauli exclusion principle.

In open (i.e., coupled to external reservoirs) solid state systems
it is more appropriate to discuss the properties of states rather then the properties of separate individual particles.
The occupation of a single-particle state is given by the single particle distribution function which is a probability to find the particle in a state under consideration.
By analogy, for a system of noninteracting particles, one can consider the probability to find several particles in some given (single-particle) states. 
This probability is a multi-particle distribution function. 
which is a diagonal element of a corresponding multi-particle density matrix.
If the particles are correlated (i.e., if the multi-particle state is not-separable) then the multi-particle distribution function does not factorize into a product of single-particle distribution functions.

In this paper we calculate the multi-particle distribution functions for  particle flows generated by a periodically driven scatterer and relate them to the noise power which can be measured in solid state structures.
We consider a mesoscopic scatterer coupled to $N_{r}$ stationary reservoirs of noninteracting electrons via single channel leads.
We will number electron states in leads with Greek letters, $\alpha,\beta,\dots$,
which include both the orbital (a lead number) and the spin indices.
We suppose that all the reservoirs are at zero temperature and have the same
Fermi energy $\mu$.
The electrons in different reservoirs are uncorrelated.
Therefore, the multi-particle incoming state is pure and separable.
After scattering by the pump the state remains pure. \cite{BTT05}
However it becomes correlated and thus non-separable. \cite{SamuelssonB04,BTT05}
Correlations between the out-going particles appear as a result of an interplay of two factors, the Pauli exclusion principle and photon-assisted scattering at a working pump.

Below, we consider, first, the effect of photon-assisted scattering which appears already in single-particle scattering. Then, in a second step, we analyze two- and multi-particle correlations.

The paper is organized as follows. In Sec.\ref{Sec2} we discuss the approximations made and analyze the single-particle distribution function for electrons scattered by the pump. In the next section, Sec.\ref{Sec3}, we calculate the two-particle correlations and their relation to the zero-frequency noise power in the case of large amplitude pumps. Then, in Sec.\ref{Sec4} we consider multi-particle correlations produced by the pump. We conclude in Sec.\ref{concl}.

\section{Single-particle scattering}
\label{Sec2}

In the presence of a scatterer oscillating with frequency $\Omega$, particles absorb or emit energy quanta while traversing the pump.
In contrast, while in the leads, they are described by the stationary Schr\"{o}dinger equation.
Thus one can classify the states of electrons (both out-going and incoming) in the leads according to a particle energy $E$ like in the stationary case.
Therefore, we will use the second quantization operators corresponding to states of particles with definite energy.
We denote the annihilation operator for incoming states $\hat{a}_{\alpha}(E)$
and the one for scattered electrons $\hat{b}_{\alpha}(E)$.
These operators are related to each other through the Floquet scattering matrix $\hat{S}_{F}$ of a periodically driven scatterer.
\cite{Shirley65,PB98,PA04}

\subsection{Adiabatic approximation}

If the pump oscillates slowly, $\Omega\to 0$, then the Floquet scattering matrix can be expressed in terms of the Fourier coefficients $\hat{S}_{m}(E)$ of the stationary scattering matrix $\hat{S}(E,\{p_{i}\})$ with time dependent parameters
$p_{i}(t+{\cal T}) = p_{i}(t)$, where ${\cal T} = 2\pi/\Omega$, as follows:
\cite{MBstrong02}
\begin{equation}
\label{Eq1}
\hat S_{F}(E,E_{m}) = \hat S_{-m}(E) + {\cal O}(\Omega).
\end{equation}
Here $E_{m} = E + m\hbar\Omega$; $m$ is an integer.
We ignore terms of order $\Omega$ and higher.
The Floquet scattering matrix couples electron states at energies shifted by one or several quanta $\hbar\Omega$.
In what follows, we will consider only electrons with energy close to the Fermi energy $\mu$.
Thus it is convenient to introduce the following notation
for energy variables: $E\equiv \epsilon_n = \epsilon + n\hbar\Omega$, where
the Floquet energy $\epsilon$ lies within the interval, $\mu < \epsilon < \mu + \hbar\Omega$.
As a consequence, the operators for out-going particles relate to the ones for incoming particles through:
\begin{equation}
\label{Eq2}
 \hat b_{\alpha}(\epsilon_n) = \sum_\beta \sum\limits_{m}
S_{\alpha\beta,n-m}\hat a_{\beta}(\epsilon_m),
\end{equation}
We evaluate the stationary scattering matrix at the Fermi energy.
Current conservation forces the Floquet scattering matrix to be unitary.
In the adiabatic approximation of Eq.(\ref{Eq1}) the unitarity reads:
\begin{equation}
\label{Eq3}
\sum\limits_{\gamma}\sum\limits_{m=-\infty}^{\infty}
S^{*}_{\gamma\alpha,m}S_{\gamma\beta,m+l} =
\sum\limits_{\gamma}\sum\limits_{m=-\infty}^{\infty}
S^{*}_{\alpha\gamma,m}S_{\beta\gamma,m+l}
= \delta_{l0}\delta_{\alpha\beta},
\end{equation}
This follows directly from the fact that the stationary scattering matrix is unitary:
$\hat S^{\dagger}\hat S = \hat S\hat S^{\dagger} = \hat {\cal I}$.
Here $\hat {\cal I}$ is a unit matrix.

At zero temperature all incoming states with $\epsilon_{m}<\mu$ ($m\le -1$) are filled while the ones with $\epsilon_{m}>\mu$ ($m\ge 0$) are empty.
The corresponding single-particle distribution function
$f_{\alpha}^{(in)}(\epsilon_m) = \langle
\hat a_{\alpha}^{\dagger}(\epsilon_m) \hat a_{\alpha}(\epsilon_m) \rangle$
is the Fermi distribution function which is the Heaviside step function at zero temperature.

\subsection{Non-equilibrium out-going particles}

The situation is different for out-going particles.
Due to interaction with an oscillating scatterer an electron can gain or lose some energy quanta $\hbar\Omega $.
Therefore, some out-going states with $\epsilon_{n}>\mu $ can be partially occupied while the ones with $\epsilon_{n}<\mu$ become partially emptied.
The probability to find an out-going particle in the state with energy $\epsilon_n$ in lead $\alpha$ is given by the single-particle distribution function: \cite{MBstrong02}
\begin{equation}
\label{Eq4}
f^{(out)}_{\alpha}(\epsilon_n) =
\langle \hat b_{\alpha}^{\dagger}(\epsilon_n)
\hat b_{\alpha}(\epsilon_n) \rangle
=\sum\limits_{\gamma}
\sum\limits_{m=-1}^{-\infty}
\left| S_{\alpha\gamma,n-m}\right|^2.
\end{equation}
Here $\langle\cdots\rangle$ denotes a quantum-statistical average. Since the $\hat{b}$-operators are expressed in terms of $\hat{a}$-operators [see, Eq.(\ref{Eq2})] we, in fact, average over (the product of) the equilibrium states of reservoirs.

To assess $f^{(out)}$ at energies far from the Fermi energy we take into account the following.
For each particular pump, there exists $n_{max}$ such that
$\hat{S}_{\pm n}\approx 0$ for $|n|\ge n_{max}$.
Then far below (above) the Fermi energy, $n\le -n_{max}$ ($n\ge +n_{max}$), all the relevant incoming states are filled (empty).
Therefore, the out-going state is filled (empty) with a unit probability.
For $n\le -n_{max}$ this follows directly from the unitarity condition  Eq.(\ref{Eq3}) taken at $l=0$  and $\alpha = \beta$.

\begin{figure}[b]
  \vspace{0mm}
  \centerline{
   \epsfxsize 8cm
   \epsffile{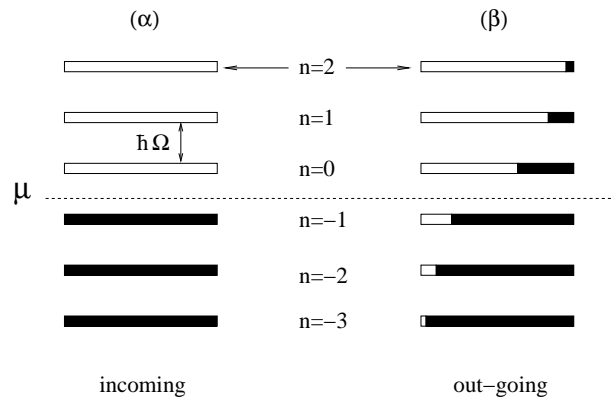}
             }
  \vspace{0mm}
  \nopagebreak
  \caption{Occupation of single particle states $\epsilon_{n} = \epsilon + n\hbar\Omega$ in incoming channels ($\alpha$) and in out-going channels ($\beta$) of a working pump at zero temperature.
The filled box area is proportional to the occupation probability.
The case shown corresponds to $n_{max}=3$, i.e. $S_{\beta\alpha,\pm |n|}\approx 0$, for $n \geq 4$.
$\mu$ is the Fermi energy;
$\Omega$ is a driving frequency.
}
\label{fig1}
\end{figure}

In contrast, in the vicinity of the Fermi energy, $\epsilon_{n}\approx\mu$
(with $|n|\le n_{max}$),
the distribution function Eq.(\ref{Eq4}) describes partially occupied states  above and below $\mu$ (see Fig.\ref{fig1}).
Since for these energies not all the relevant
[i.e., the states with energy $\epsilon_m$ for which $-n_{max} \le n-m \le +n_{max}$] incoming states are filled (empty).
The incoming states corresponding to $n-n_{max}\le m\le -1$ are filled while those with $0\le m \le n+n_{max}$ are empty.
Close to the Fermi energy particles in $n_{max}N_{r}$ incoming channels are scattered into $2n_{max}N_{r}$ available out-going channels.
Since the number of available out-going channels is larger then the number of
incoming channels, the appearance of a particle in some out-going state is
a random process.
This results in a partial occupation of the corresponding single-particle states and leads to the appearance of particle-particle correlations.

Notice, the pump is able to correlate only those out-going particles which have energies shifted by one or several quanta $\hbar\Omega$. While the particles having different Floquet energies, $\epsilon\neq\epsilon^{\prime}$, remain uncorrelated like the incoming particles.

To investigate the correlations we calculate the joint probability to find several out-going channels occupied and compare it to the product of occupation probabilities of individual channels.
The single-channel occupation probability is the one-particle distribution function. In analogy we calculate multichannel occupation probabilities which are multi-particle distribution functions.
First we consider the two-particle probability.

\section{Two-particle scattering}
\label{Sec3}

To calculate the joint probability to register one out-going particle with energy $\epsilon_{n}$ at lead $\alpha$ and other one with energy
$\epsilon_{m}$ at lead $\beta$ we introduce the two-particle distribution function
$f^{(out)}_{\alpha,\beta}(\epsilon_{n},\epsilon_{m})$
for out-going particles.
This distribution function is the quantum-statistical average
\begin{equation}
\label{Eq5}
f^{(out)}_{\alpha,\beta}(\epsilon_{n},\epsilon_{m}) = \langle
\hat B_{\epsilon_{n},\epsilon_{m}}^{\dagger}(\epsilon_{n},\epsilon_{m})
\hat B_{\alpha,\beta}(\epsilon_{n},\epsilon_{m})\rangle,
\end{equation}
of the two-particle operator
\begin{equation}
\label{Eq5_1}
\hat B_{\alpha,\beta}(\epsilon_{n},\epsilon_{m}) =
\hat b_{\alpha}(\epsilon_{n})\hat b_{\beta}(\epsilon_{m}).
\end{equation}
Using Eqs.(\ref{Eq2}) and (\ref{Eq4}) the distribution function can be expressed as a sum of two contributions,
\begin{equation}
\label{Eq6_0}
f^{(out)}_{\alpha,\beta}(\epsilon_{n},\epsilon_{m}) =
f^{(out)}_{\alpha}(\epsilon_{n}) f^{(out)}_{\beta}(\epsilon_{m})
+ \delta f^{(out)}_{\alpha,\beta}(\epsilon_{n},\epsilon_{m}),
\end{equation}
one which factorizes and an irreducible part, a two-particle correlation function,
\begin{equation}
\label{Eq6}
\begin{array}{l}
\delta f^{(out)}_{\alpha,\beta}(\epsilon_{n},\epsilon_{m}) =
-\left| K_{\alpha\beta}(\epsilon_{n},\epsilon_{m})\right|^2, \\
\ \\
K_{\alpha\beta}(\epsilon_{n},\epsilon_{m}) =
\sum\limits_{\gamma}
\sum\limits_{p=-1}^{-\infty}
S_{\alpha\gamma,n-p}
S_{\beta\gamma,m-p}^{*},
\end{array}
\end{equation}
where
$K_{\alpha\beta}(\epsilon_{n},\epsilon_{m}) = \langle
\hat b_{\beta}^{\dagger}(\epsilon_{m}) 
\hat b_{\alpha}(\epsilon_{n})\rangle$
is a matrix element of the pair correlator. \cite{BTT05}

If $K_{\alpha\beta}(\epsilon_{n},\epsilon_{m})$ is non-vanishing the two-particle probability does not factorize into the product of single-particle ones and we conclude that the particles are correlated.

The electron-electron correlation function
 $\delta f^{(out)}_{\alpha,\beta}(\epsilon_{n},\epsilon_{m})$
characterizes correlations between the out-going particles.
Since $\delta f^{(out)}_{\alpha,\beta}$ is the (negative of a) square of some quantity we can interpret it as a probability of some scattering process which is responsible for two-particle correlations. This process consists of the creation and scattering of electron-hole pairs.

\subsection{Two-particle correlations: an electron-hole pair view}

The matrix element $K_{\alpha\beta}(\epsilon_{n},\epsilon_{m})$ of a pair correlator is a sum of particular amplitudes
\begin{equation}
\label{Eq7}
A_{\alpha,\beta}^{(\gamma;p)}(\epsilon_{n},\epsilon_{m}) =
S_{\alpha\gamma,n-p}
S_{\beta\gamma,m-p}^{*},
\end{equation}
which can be related to the excitation of an electron-hole pair from
the filled state with energy $\epsilon_{p}$ incident from reservoir $\gamma$.
Formally one can consider the hole (an empty state) as a particle
with corresponding second quantized operators
$\hat{a}^{(h)} = \hat{a}^{\dagger}, \hat{b}^{(h)} = \hat{b}^{\dagger}$.
Therefore, scattering of holes is described by the adjoint scattering matrix $\hat{S}^{\dagger}$.
We define holes for any energies, above and below the Fermi energy.
The hole can be scattered by the pump like an electron.
Both electron and hole can emit/absorb some energy quanta $\hbar\Omega$.
Therefore, the amplitude Eq.(\ref{Eq7}) corresponds to the process with a final state in which an electron with energy $n\hbar\Omega$
is scattered into lead $\alpha$ and
a hole with energy $m\hbar\Omega$ is scattered into lead $\beta$.
To get the whole amplitude
$K_{\alpha\beta}(\epsilon_{n},\epsilon_{m})$
it is necessary to sum over all the indistinguishable processes
leading to a given final state.
Different processes are possible due to the presence of many leads (index $\gamma$) from which the pair is born and due to different filled energy channels (index $p$).
Alternatively the quantity
$\left(A_{\alpha,\beta}^{(\gamma;p)}(\epsilon_{n},\epsilon_{m})\right)^{*}$
is an amplitude for electron and hole to be scattered to leads $\beta$ and $\alpha$, respectively.
Therefore, the two-particle correlation function
$\delta f^{(out)}_{\alpha,\beta}(\epsilon_{n},\epsilon_{m})$, Eq.(\ref{Eq6}),
is the probability to create an electron-hole pair and
to emit its constituents (an electron and a hole) into two chosen states.
Notice, the shot noise generated by the pump is due to the same processes. \cite{MB02}

We remark that the electron-hole pair shot noise generated by a microwave field applied to the contacts of the sample has recently been measured. \cite{reydellet} In this experiment there is no applied dc-voltage and hence no dc-current. Refs.~\onlinecite{rychkov} and \onlinecite{polianski} propose experiments with multiple microwave fields of the same frequency but with possible phase-lags
applied to different contacts of the sample. In contrast to the discussion 
provided here, since the microwave fields are applied to the contacts, the scattering
matrix is (apart from self-consistent effects \cite{polianski}) stationary.

One can ask why the correlations between two electrons depend on
processes involving electron-hole pairs.
That is due to the fact that in each state the (operators for the)
numbers of electrons and holes sum up to unity.
Therefore, the fluctuation of the number of electrons and holes are not independent from each other.
As a result the various two-particle correlation functions involving electrons and/or holes
are related to each other in a simple way:
\begin{equation}
\label{Eq8}
\begin{array}{c}
\delta f^{(out)}_{\alpha,\beta}(\epsilon_{n},\epsilon_{m})
= - \delta f^{(out)}_{\alpha,\beta}(0_{\epsilon_{n}},\epsilon_{m}) \\
\ \\
= - \delta f^{(out)}_{\alpha,\beta}(\epsilon_{n},0_{\epsilon_{m}})
= \delta f^{(out)}_{\alpha,\beta}(0_{\epsilon_{n}},0_{\epsilon_{m}}).
\end{array}
\end{equation}
Here $0_{\epsilon_{n}}$ means that there is no particle (there is a hole) in the state $\epsilon_{n}$.
Note, we define an electron-hole distribution function, say
$f^{(out)}_{\alpha,\beta}(\epsilon_{n},0_{\epsilon_{m}}) $,
by Eq.(\ref{Eq5})
with an electron operator $\hat b_{\beta}(\epsilon_{m})$ being replaced by
the corresponding hole operator
$\hat b^{(h)}_{\beta}(\epsilon_{m}) =  \hat b^{\dagger}_{\beta}(\epsilon_{m})$.
The various two-particle distribution functions sum up to unity
\begin{equation}
\label{Eq9}
\begin{array}{c}
f^{(out)}_{\alpha,\beta}(\epsilon_{n},\epsilon_{m})
+ f^{(out)}_{\alpha,\beta}(0_{\epsilon_{n}},\epsilon_{m})  \\
\ \\
+ f^{(out)}_{\alpha,\beta}(\epsilon_{n},0_{\epsilon_{m}})
+ f^{(out)}_{\alpha,\beta}(0_{\epsilon_{n}},0_{\epsilon_{m}}) = 1.
\end{array}
\end{equation}
This normalization condition justifies the point of view that
$f^{(out)}_{\alpha,\beta}$ is a two-particle joint probability.
In addition Eq.(\ref{Eq9}) shows that the electron-hole processes affect
the electron-electron probability.

Note, that the whole electron-hole scattering amplitude
$K_{\alpha\beta}(\epsilon_{n},\epsilon_{m})$
is a sum of a number of two-particle (electron-hole) photon-assisted scattering amplitudes
$A_{\alpha,\beta}^{(\gamma;p)}(\epsilon_{n},\epsilon_{m})$
which interfere between themselves.
When all the relevant incoming states are filled then owing to current conservation [i.e., to a unitarity of scattering, Eq.(\ref{Eq3})] the electron-hole amplitudes cancel each other and the correlations disappear.
This holds for particles with energies far below the Fermi energy.
That is quite similar to the stationary case when the unitary scattering process
does not produce additional correlations if incoming states are not correlated.
\cite{KSBK02,XbWang02}
In contrast for the particles with energies close to the Fermi energy not all the relevant incoming state are filled and the correlations appear.

\subsection{Pauli exclusion principle and two-particle joint probability}

The correlations under consideration are quantum-mechanical because they originate from the Pauli exclusion principle.
According to this principle two particles can not be simultaneously scattered into the same out-going state, and any incoming state can not be a source for two particles.
Formally the Pauli principle leads to the existence of two amplitudes
(direct and exchange) describing scattering of two fermions.
In our case this becomes evident if we express the two-particle distribution
function in terms of Slater determinants:
\begin{equation}
\label{Eq10}
\begin{array}{c}
f^{(out)}_{\alpha,\beta}(\epsilon_{n},\epsilon_{m}) =
\frac{1}{2}\sum\limits_{\gamma} \sum\limits_{\delta}
\sum\limits_{p=-1}^{-\infty}\sum\limits_{q=-1}^{-\infty}
\left| det \hat{M}^{(2)} \right|^2, \\
\ \\
\hat{M}^{(2)} = \left(
\begin{array}{cc}
S_{\alpha\gamma,n-p} & S_{\alpha\delta,n-q} \\
S_{\beta\gamma,m-p} &  S_{\beta\delta,m-q}
\end{array}
\right).
\end{array}
\end{equation}
Thus, the joint probability
to measure two particles in some given out-going states
$|\alpha;\epsilon_{n}\rangle$ and $|\beta;\epsilon_{m}\rangle$
is a sum of various probabilities describing photon-assisted scattering
of two electrons.
Each such scattering is described by the Slater determinant whose elements are
photon-assisted single-particle scattering amplitudes.
The sum of squared direct/exchange two-particle amplitudes gives the product
of two single-particle distribution functions
$f^{(out)}_{\alpha}(\epsilon_n)f^{(out)}_{\beta}(\epsilon_m)$
(it is a joint probability for uncorrelated particles).
While the interference between direct and exchange amplitudes is
responsible for electron-electron correlations.

\subsection{Current noise and two-particle correlations}

The quantity
$\delta f^{(out)}_{\alpha,\beta}(\epsilon_{n},\epsilon_{m})$
characterizes how much two particles in the states
$|\alpha;\epsilon_{n}\rangle$ and $|\beta;\epsilon_{m}\rangle$
are correlated between themselves.
To measure these correlations
(as well as to measure single-particle distribution functions)
one needs to use energy filters with resolution better then $\hbar\Omega$.
Apparently this is a challenge for an experimental realization especially
in the adiabatic regime, $\Omega\to 0$.
Therefore, it is useful to relate what we calculated to what can be easier
measured experimentally.
In mesoscopic systems the zero frequency noise power is an observable which
characterizes the correlations between the particles.
\cite{Buttiker90,SB05,Beenakker05}

The zero frequency noise power,
\begin{equation}
\label{Eq11}
{\bf P}_{\alpha\beta} = \frac{1}{2} \int\limits_{0}^{\cal T} \frac{dt}{{\cal T}}
\int\limits_{-\infty}^{\infty} d\tau
\langle
\Delta\hat I_{\alpha}(t)\Delta\hat
I_{\beta}(t + \tau) +
\Delta\hat I_{\beta}(t+ \tau) 
\Delta\hat I_{\alpha}(t) \rangle,
\end{equation}
produced by the pump was calculated in
Ref.~\onlinecite{MB02}.
Here $\hat I_{\alpha}(t)$ is the current operator in
lead $\alpha$ and $\Delta\hat I_{\alpha}(t) = \hat I_{\alpha}(t) -
\langle\hat I_{\alpha}(t)\rangle$.
In terms of
$\delta f^{(out)}_{\alpha,\beta}(\epsilon_{n},\epsilon_{m})$
the zero-temperature current-current cross-correlator
($\alpha\ne\beta$) reads
\begin{equation}
\label{Eq12}
{\bf P}_{\alpha\beta} = e^2\frac{\Omega}{2\pi}
\sum\limits_{n=-\infty}^{\infty}
\sum\limits_{m=-\infty}^{\infty}
\delta f^{(out)}_{\alpha,\beta}(\epsilon_{n},\epsilon_{m}),
\quad \alpha\ne\beta.
\end{equation}
The auto-correlator ${\bf P}_{\alpha\alpha}$ can be calculated
using the conservation law $\sum_{\alpha}{\bf P}_{\alpha\beta} = 0$.
Notice the factor $\Omega/2\pi$ counts all the statistically independent
sets of states corresponding to the different Floquet energies $0<\epsilon-\mu<\hbar\Omega$.

We see that the zero-frequency noise power depends on pair correlations between  the out-going particles in different energy channels exiting the pump through two chosen leads $\alpha$ and $\beta$.
Therefore, it can be used as a measure of two-particle correlations produced by the pump. That agrees with conclusions made in Ref.~\onlinecite{SamuelssonB04}.
However the pump produces the multi-particle correlations as well.

\section{Multi-particle scattering}
\label{Sec4}

To show this we introduce an $N$-particle operator
\begin{equation}
\label{Eq13}
\hat B_{\alpha_1,\cdots,\alpha_N}(\epsilon_{n_{1}},\cdots,\epsilon_{n_{N}}) =
\prod\limits_{j=1}^{N} \hat b_{\alpha_{j}}(\epsilon_{n_{j}}),
\end{equation}
and the corresponding $N$-particle distribution function
$ f^{(out)}_{\alpha_1,\cdots,\alpha_N} =  \langle
B_{\alpha_1,\cdots,\alpha_N}^{\dagger} B_{\alpha_1,\cdots,\alpha_N}\rangle$,
which can be represented in terms of $N\times N$ Slater determinants
\begin{equation}
\label{Eq14}
\begin{array}{l}
f^{(out)}_{\alpha_1,\cdots,\alpha_N}(\epsilon_{n_{1}},\cdots,\epsilon_{n_{N}})
 = \frac{1}{N!}
\sum\limits_{\gamma_1...\gamma_N}
\sum\limits_{p_1...p_N=-1}^{-\infty}
\left| det \hat{M}^{(N)} \right|^2, \\
M_{ij}^{(N)} = S_{\alpha_{i}\gamma_{j},n_{i}-p_{j}}, \quad i,j = 1,\cdots,N.
\end{array}
\end{equation}
The $N$-particle correlations can be defined as follows.
Let us subtract from the $N$-particle distribution function
various products of distribution functions and correlations functions corresponding to a smaller number of particles.
Then the remaining $N$-particle correlation function
$\delta f^{(out)}_{\alpha_1,\cdots,\alpha_N}(\epsilon_{n_{1}},\cdots,
\epsilon_{n_{N}})$
will describe the correlations shared by $N$ scattered particles.
In terms of elements of the pair correlator it reads as follows:
\begin{equation}
\label{Eq15}
\begin{array}{c}
\delta f^{(out)}_{\alpha_1,\cdots,\alpha_N}(\epsilon_{n_{1}},\cdots,
\epsilon_{n_{N}}) = (-1)^{N-1} \sum\limits_{P_{N}^{\prime}} \\
\times
K_{\alpha_{r_{1}}\alpha_{r_{N}}}(\epsilon_{n_{r_{1}}},
\epsilon_{n_{r_{N}}})
 \prod\limits_{j=1}^{N-1}
K_{\alpha_{r_{j+1}}\alpha_{r_{j}}}(\epsilon_{n_{r_{j+1}}},
\epsilon_{n_{r_{j}}}).
\end{array}
\end{equation}
Here the sum runs over the set of all non-equivalent
permutations $P_{N}^{\prime} = (r_{1},r_2,\cdots,r_N)$ of integers from $1$ to $N$.
Non-equivalent permutations of integers are those that can not be obtained via a cyclic permutation of each other.

Notice, the multi-particle correlation function, Eq.(\ref{Eq15}), for $N>2$ can be either positive or negative unlike the two-particle correlation function, Eq.(\ref{Eq6}), which is definitely negative.

\subsection{Three-particle correlations}

\begin{figure}[t]
  \vspace{0mm}
    {
   \epsfxsize 10cm
   \epsffile{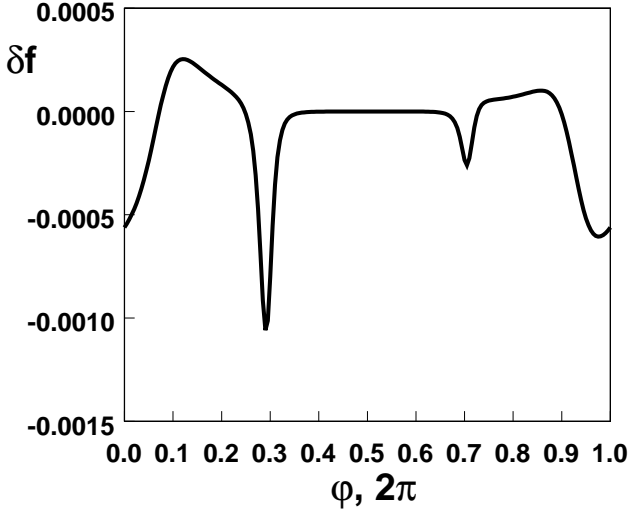}
             }
  \vspace{0mm}
  \nopagebreak
  \caption{Three-particle correlation function
$\delta f^{(out)}_{L,L,R}(\epsilon_{n_{0}}, \epsilon_{n_{1}},\epsilon_{n_{-1}})$
as a function of the phase lag $\varphi$.
The scattering potential is:
$V(x,t) = \delta(x+a/2)[V_0 + 2V_1\cos(\Omega t)] + \delta(x-a/2)[V_0+2V_1\cos(\Omega t + \varphi)]$
The parameters are:
$a=100\pi$; $V_0=20$; $V_1=10$;
and the Fermi energy $\mu=1.0197$.
We use the units $2m_e =\hbar = e =1$, where $m_e$ is the mass of an electron.
}
\label{fig2}
\end{figure}

As an example we consider the three-particle distribution function in detail (for brevity we suppressed corresponding energy arguments):
\begin{equation}
\label{Eq16}
\begin{array}{c}
f^{(out)}_{\alpha,\beta,\gamma} =
f^{(out)}_{\alpha} f^{(out)}_{\beta} f^{(out)}_{\gamma}
+ f^{(out)}_{\alpha} \delta f^{(out)}_{\beta,\gamma} \\
\ \\
+  f^{(out)}_{\beta} \delta f^{(out)}_{\alpha,\gamma}
+  f^{(out)}_{\gamma} \delta f^{(out)}_{\alpha,\beta}
+  \delta f^{(out)}_{\alpha,\beta,\gamma},\\
\ \\
\delta f^{(out)}_{\alpha,\beta,\gamma}(\epsilon_{n_{\alpha}},
\epsilon_{n_{\beta}},\epsilon_{n_{\gamma}}) = 2 Re\left[
\sum\limits_{\chi} \sum\limits_{r=-1}^{-\infty}
S_{\alpha\chi,n_{\alpha}-r}
S_{\gamma\chi,n_{\gamma}-r}^{*}
\right. \\
\left.
\times 
\sum\limits_{\delta} \sum\limits_{p=-1}^{-\infty}
S_{\beta\delta,n_{\beta}-p} 
S_{\alpha\delta,n_{\alpha}-p}^{*}
\sum\limits_{\varphi} \sum\limits_{q=-1}^{-\infty}
S_{\gamma\varphi,n_{\gamma}-q}
S_{\beta\varphi,n_{\beta}-q}^{*}
\right],
\end{array}
\end{equation}
where the single-particle distribution function
$f^{(out)}_{\alpha}$
and the two-particle correlation function
$\delta f^{(out)}_{\alpha,\beta}$
are defined in Eq.(\ref{Eq4}) and Eq.(\ref{Eq6}), respectively.
In the stationary case
a three-particle correlation function is positive,
$\delta f^{(out)}_{\alpha,\beta,\gamma}(\epsilon_{n_{\alpha}},
\epsilon_{n_{\beta}},\epsilon_{n_{\gamma}})=
2\delta_{\alpha,\beta}\delta_{\beta,\gamma}
\delta_{n_\alpha,n_\beta}\delta_{n_\beta,n_\gamma}$.
In contrast in the driven case the sign can be either positive or negative.

In Fig.\ref{fig2} we give the three-particle correlation function
$\delta f^{(out)}_{L,L,R}(\epsilon_{n_{0}}, \epsilon_{n_{1}},\epsilon_{n_{-1}})$
for a two-terminal 1D scatterer composed by two oscillating 
$\delta$-function potentials placed a distance $a$ from one another.
The indices $L$ and $R$ stand for the left and right terminals, respectively.
The variation of the phase lag $\varphi$ 
between oscillating potentials
changes the value and the sign of the three-particle correlation function.

\subsection{Generating function}

Formally one can represent the N-particle correlation function, Eq.(\ref{Eq15}), as the N-th order derivative of some generating function dependent on the pair correlator $\hat K$. 
To this end we introduce a diagonal matrix 
$\hat\Lambda = {\rm diag}(\lambda_\rho)$
of auxiliary fields $\lambda_\rho$., where 
$\rho = 1,2,\cdots,N_{max}$ with $N_{max}=(2n_{max}+1)N_{r}$.
The index $\rho\equiv(\alpha,n)$ includes both the lead index $\alpha = 1,2,\cdots,N_{r}$ and the energy channel index $n=0,\pm 1,\cdots,\pm n_{max}$.
Then the N-particle correlation function can be calculated as
\begin{equation}
\label{Eq19}
\left.
\delta f^{(out)}_{\rho_1,\cdots,\rho_N} =
\frac{\partial^{N}{\rm Tr}
\ln\left( \hat I + \hat\Lambda\hat K\right)
}{\partial\lambda_{\rho_1}\cdots
\partial\lambda_{\rho_N}}\right|_{\lambda_{\rho}=0}.
\end{equation}
Here $\hat I$ is a unit matrix of dimension $N_{max}\times N_{max}$.
For $N=1$ we obtain the single-particle distribution function, Eq.(\ref{Eq4}).

To prove that Eq.(\ref{Eq19}) leads to Eq.(\ref{Eq15})
it is necessary to use the Taylor expansion of Eq.(\ref{Eq19})
in powers of $\hat\Lambda$.
Only the N-th term of this expansion does contribute to the quantity of interest.

We emphasize that Eq.(\ref{Eq19}) determines a multi-particle correlation function which is a different quantity then transferred charge which is usually discussed. \cite{ILL97}
The important difference is that here we are interesting in energy-resolved correlation characteristics of pumped particles rather then in statistical properties of the whole current carried by the particles at all the energies. 

\subsection{Higher order current cumulants}

The multi-particle correlations generated by the pump can be related to higher order current cumulants in full analogy with how the two-particle correlations are related to the zero-frequency noise power, Eq.(\ref{Eq12}). 

We consider the zero-frequency Fourier transform of the Nth-order current correlation function symmetrized in lead indices 
\begin{equation}
\label{Eq20}
{\bf P}_{\alpha_1\cdots\alpha_N} 
= \frac{1}{N!} \int\limits_{0}^{\cal T} \frac{dt}{{\cal T}}
\int\limits_{-\infty}^{\infty} d\tau_2\cdots d\tau_N 
\sum\limits_{P_{N}}
\langle
\prod\limits_{i=1}^{N}\Delta\hat I_{\alpha_{r_i}}(t+\tau_{r_i})
\rangle.
\end{equation}
Here the sum runs over the set of all the permutations
$P_{N}=(r_1,\cdots,r_N)$ of integers from 1 to N.
We suppose $\tau_1=0$.

At zero temperature the current cross-correlator 
($\alpha_1\neq\cdots\neq\alpha_N$) can be expressed in terms of the N-particle correlation functions for outgoing particles as follows:
\begin{equation}
\label{Eq21}
{\bf P}_{\alpha_1\cdots\alpha_N} 
= \frac{e^N\Omega}{2\pi}
\sum\limits_{n_1=-\infty}^{\infty}\cdots
\sum\limits_{n_N=-\infty}^{\infty}
\delta f^{(out)}_{\alpha_1,\cdots,\alpha_N}(\epsilon_{n_{1}},\cdots,
\epsilon_{n_{N}})
\end{equation}
Such a relation follows straightforwardly from two observations. First, at zero temperature only the currents carried by the outgoing particles contribute to the current correlation function. And, second, the operators for outgoing particles in the expression for
the current cross-correlation function can be arranged into the same order as in the expression for the N-particle correlation function.

Therefore, the N-particle correlations generated by the pump can be experimentally probed via the Nth-order cross-correlator of currents flowing into the leads attached to the pump.

\section{Discussion and Conclusion}
\label{concl}

We have shown that an adiabatic quantum pump generates multi-particle
correlations which originate from the Pauli exclusion principle and
appear in photon-assisted scattering channels. 
Formally the pump produces multi-particle correlations up to an infinite order ($N\to\infty$). This is because the pump can excite all the incoming particles.
In fact, as we mentioned already, this number can be bounded by $n_{max}N_r$.

The multi-particle correlations of all orders are important to characterize the whole outgoing state produced by the pump. 
For instance, one projects the whole state onto a state with an exact number of excited particles. 
Such a projected state is a multi-particle state, therefore, its statistical-correlation properties depend on multi-particle correlations as well.
In Ref.~\onlinecite{BTT05} the projected electron-hole pair state produced by the pump was considered and the electron-hole entanglement entropy was calculated. Since the entanglement entropy depends on multi-particle correlations, it, in a general case, can not be related to charge noise representing only two-particle correlations, Eq.(\ref{Eq12}).
However in some particular cases the pump produces effectively only two-particle correlations.
This is the case for a weak amplitude ($n_{max}=1$), spin-independent pump with two single-channel leads ($N_r=2$). In this case entanglement entropy can be related to charge noise. \cite{SamuelssonB04,BTT05}

The multi-particle correlations  Eq.(\ref{Eq15}), produced by the pump depend on the elements of the Floquet scattering matrix.
These elements are amplitudes of scattering with absorption or emission of one or several modulation quanta $\hbar\Omega$.
The maximum number $n_{max}$ of energy quanta which can be absorbed/emitted during the scattering process define the maximum order of multi-particle correlations produced by the pump.
The number $n_{max}$ depends strongly on the strength of driving.
At weak driving $n_{max}=1$ while at strong driving $n_{max}\gg 1$.
On the other hand the value of a multi-particle correlation function can be changed by changing the parameters of driving, for instance, the phase difference between two driving parameters.
Therefore, the order and the amount of correlations generated by a pump can be simply manipulated by changing the parameters of a drive.

\begin{acknowledgments}

M.M. appreciates the warm hospitality of the Department of Theoretical Physics of the University of Geneva, where part of this work was done.
This work was supported by the Swiss National Science Foundation.

\end{acknowledgments}

\end{document}